\documentclass[twoside]{dis08}
\usepackage[latin1]{inputenc}
\usepackage[dvips]{graphicx,epsfig,color}
\usepackage{wrapfig,rotating}
\usepackage{amssymb,amsmath,array}

\newcommand {\pom} {I\!\!P}

\newcommand {\reg} {I\!\!R}

\newcommand{\st}{$\sigma_{\mathrm {tot}}(\gamma p)$ }
\def\ap#1#2#3   {{Ann. Phys. (NY)} {\bf#1} (#2) #3}   
\def\apj#1#2#3  {{Astrophys. J.} {\bf#1} (#2) #3} 
\def\apjl#1#2#3 {{Astrophys. J. Lett.} {\bf#1} (#2) #3}
\def\app#1#2#3  {{Acta. Phys. Pol.} {\bf#1} (#2) #3}
\def\ar#1#2#3   {{Ann. Rev. Nucl. Part. Sci.} {\bf#1} (#2) #3}
\def\cpc#1#2#3  {{Computer Phys. Comm.} {\bf#1} (#2) #3}
\def\epj#1#2#3  {{Eur. Phys. J.} {\bf#1} (#2) #3}
\def\err#1#2#3  {{Erratum} {\bf#1} (#2) #3}
\def\ib#1#2#3   {{ibid.} {\bf#1} (#2) #3}
\def\jmp#1#2#3  {{J. Math. Phys.} {\bf#1} (#2) #3}
\def\ijmp#1#2#3 {{Int. J. Mod. Phys.} {\bf#1} (#2) #3}
\def\jetp#1#2#3 {{JETP Lett.} {\bf#1} (#2) #3}
\def\jpg#1#2#3  {{J. Phys. G.} {\bf#1} (#2) #3}
\def\mpl#1#2#3  {{Mod. Phys. Lett.} {\bf#1} (#2) #3}
\def\nat#1#2#3  {{Nature (London)} {\bf#1} (#2) #3}
\def\nc#1#2#3   {{Nuovo Cim.} {\bf#1} (#2) #3}
\def\nim#1#2#3  {{Nucl. Instrum. Meth.} {\bf#1} (#2) #3}
\def\np#1#2#3   {{Nucl. Phys.} {\bf#1} (#2) #3}
\def\pcps#1#2#3 {{Proc. Cam. Phil. Soc.} {\bf#1} (#2) #3}
\def\pl#1#2#3   {{Phys. Lett.} {\bf#1} (#2) #3}
\def\prep#1#2#3 {{Phys. Rep.} {\bf#1} (#2) #3}
\def\prev#1#2#3 {{Phys. Rev.} {\bf#1} (#2) #3}
\def\prl#1#2#3  {{Phys. Rev. Lett.} {\bf#1} (#2) #3}
\def\prs#1#2#3  {{Proc. Roy. Soc.} {\bf#1} (#2) #3}
\def\ptp#1#2#3  {{Prog. Th. Phys.} {\bf#1} (#2) #3}
\def\ps#1#2#3   {{Physica Scripta} {\bf#1} (#2) #3}
\def\rmp#1#2#3  {{Rev. Mod. Phys.} {\bf#1} (#2) #3}
\def\rpp#1#2#3  {{Rep. Prog. Phys.} {\bf#1} (#2) #3}
\def\sjnp#1#2#3 {{Sov. J. Nucl. Phys.} {\bf#1} (#2) #3}
\def\spj#1#2#3  {{Sov. Phys. JEPT} {\bf#1} (#2) #3}
\def\spu#1#2#3  {{Sov. Phys.-Usp.} {\bf#1} (#2) #3}
\def\zp#1#2#3   {{Zeit. Phys.} {\bf#1} (#2) #3}
\begin{document}
\title{Energy dependence of $\sigma_{tot}(\gamma p)$ at HERA}

\author{Aharon Levy$^{1,2,}$\thanks{Partly supported by the Israel Science 
Foundation (ISF).}\\
 \ \ on behalf of the ZEUS collaboration
%
%
\vspace{.3cm}\\
%
1- Max-Planck-Institute \\
F\"ohringer Ring 6, 80805 Munich, Germany
%
\vspace{.1cm}\\
2- School of Physics and Astronomy \\
Tel Aviv University, 69978 Tel Aviv, Israel\\
}

\maketitle

\begin{abstract}
The energy dependence of the total photon-proton cross-section is 
determined from data collected with the ZEUS detector at HERA with two 
different proton beam energies.
\end{abstract}

\section{Introduction}

Donnachie and Landshoff (DL)~\cite{DL} showed that the energy dependence of
all hadron-hadron total cross sections can be described by a simple 
Regge motivated form,
\vspace{-2mm}
\begin{equation}
\sigma_{\rm tot} = A \cdot (W^2)^{\alpha_{\pom}(0)-1} + B \cdot (W^2)^{\alpha_{\reg}(0)-1} \, ,
\label{totxsec}
\vspace{-1mm}
\end{equation}
where $A$ and $B$ are process dependent
constants, $W$ is the hadron-hadron center-of-mass energy, and
$\alpha_{\pom}(0)$ ($\alpha_{\reg}(0)$) is the Pomeron (Reggeon) 
trajectory intercept.

The \st dependence on $W$ is particularly interesting 
because of the nature of the photon, which is known to exhibit properties 
of both a point-like particle (direct photon) and a hadron-like 
state (resolved photon). At the $ep$ collider HERA,  
\st can be extracted from $ep$ scattering at very low momentum 
transferred squared at the electron vertex, $Q^2\simeq 0$~GeV$^2$.   

The first measurements of the total $\gamma p$ cross section at
HERA~\cite{ZEUSpaper92,H1paper93} showed that the total
photoproduction cross section has a $W$ dependence similar to that of
hadron-hadron reactions. Further measurements of \st at
HERA~\cite{ZEUSpaper94,H1paper95,ZEUSpaper02} have reduced its
statistical errors but the systematic uncertainties remained too large
for a precise determination of the $W$ dependence of the cross
section. The original fits of DL gave
$\epsilon\equiv\alpha_{\pom}(0)-1=0.0808$ and no uncertainties were
determined. Cudell et al.~\cite{Cudell}, using a data
set~\cite{dataset} that contained 2747 data points for total cross
sections and 303 data points for real part of hadronic amplitudes,
determined $\epsilon = 0.093 \pm 0.003$. However, only very few points
were present for the highest center-of-mass energies. The data were
from different experiments and had a large spread.  Furthermore, the
value of the Pomeron intercept comes out strongly correlated with that
of the Reggeon trajectories. In another evaluation~\cite{cudell2}, the
authors give the range of 0.07 - 0.10 as acceptable values for
$\epsilon$.

Recently, just before the shut-down of the HERA collider, runs with
different proton energies were taken, at constant positron energy.
This opened up the possibility to determine precisely the power of the
$W$ dependence by measuring the ratios of cross sections in one
experiment, thus having many of the systematic uncertainties canceling
out.

The difficulty in measuring total cross section, $\sigma$, in a
collider environment originates from the limited acceptance of
collider detectors for certain class of processes, in particular for
elastic and diffractive scattering, where the final state particles
are likely to disappear down the beam-pipe.  The determination of the
acceptance relies on Monte Carlo simulation of the physics and of the
detector. The simulation of the physics is subject to many
uncertainties which then impact on the systematic uncertainty on cross
section measurement.  For the energy dependence of $\sigma$, the
impact of these uncertainties as well as of the geometrical
uncertainties can be minimized by studying the ratio $r$ of cross
sections probed at different $W$ values.

\noindent
Assuming $\sigma \sim W^\delta$~\cite{ZEUSpaper02},
\begin{equation}
r=\frac{\sigma(W_1)}{\sigma(W_2)}=\left(\frac{W_1}{W_2}\right)^\delta \, .
\end{equation}
Experimentally,
\begin{equation}
\sigma = \frac{N}{A\cdot \cal{L}} \, ,
\end{equation}
where $A$, $\cal L$ and $N$ are the acceptance, luminosity and number
of measured events, respectively, and therefore
\begin{equation}
r=\frac{N_1}{N_2}\cdot \frac{A_2}{A_1} \cdot \frac{{\cal L}_{2}}{{\cal L}_{1}} \, ,
\label{acc}
\end{equation}
where the index 1(2) denotes measurements performed at $W_1$
($W_2$). The acceptance for $\gamma p$ events at HERA depends mainly
on the detector infrastructure in the electron (rear) direction. If
the change in the $W$ value results from changing the proton energy,
the acceptance is likely to remain the same independent of $W$ and
the ratio of acceptances will drop out in Eq.~(\ref{acc}).

The number of measured events is from $e^+ p$ interactions. In order
to convert the $ep$ to $\gamma p$ cross section, one uses the
following relation:
\begin{equation}
\frac{d\sigma^{e^{+}p}(y)}{dy}=\frac{\alpha}{2\pi}\left[\frac{1+(1-y)^2}{y}ln\frac
{Q^2_{max}}{Q^2_{min}}-2\frac{(1-y)}{y}(1-\frac{Q^2_{min}}{Q^2_{max}})\right]
\sigma^{\gamma p}_{tot}(y)\, ,
\label{gammapxsec}
\end{equation}
where the term multiplying the $\gamma p$ cross section is known as
the flux factor $f$. In Eq.~(\ref{gammapxsec}), $\alpha$ is the
electromagnetic coupling constant, $y$ is the energy fraction
transferred by the positron to the photon in the proton rest system,
$Q^2_{min}={m_e}^2\frac{y^2}{1-y}$ (here $m_{e}$ is the mass of the
electron) and the highest measured $Q^2$ is $Q^2_{max}$
($\sim$~0.01~GeV$^2$). The flux $f$ is a function of $y$ and needs to
be integrated over the measured $y$ range for obtaining the total
$\gamma p$ cross section.

In the final days of HERA running, a special photoproduction trigger
was implemented for runs with proton energy set at the nominal value
of 920 GeV (high-energy run, HER) and at proton energies lowered to
575 (medium-energy run, MER) and 460 GeV (low-energy run, LER). The
analysis presented here attempts to extract the value of $\delta$
based on the combination of the HER and LER data.

\section{Experimental details}

\begin{figure}[t]
\includegraphics[scale=0.6,angle=270,clip]{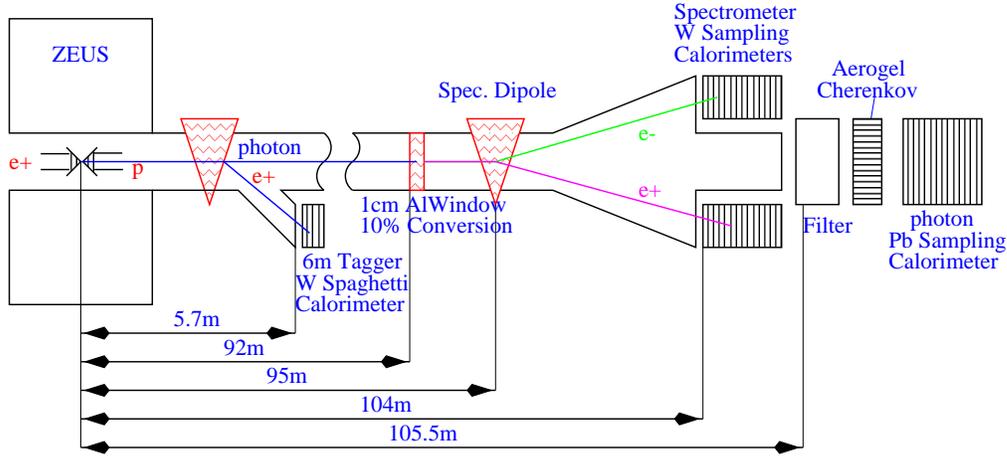}
\caption{ Schematic layout of the ZEUS detector with the six meter tagger
and the components of the luminosity system, together with their
distance from the interaction point.}
\label{fig:setup}
\end{figure}
The setup of the experiment is shown in Fig.~\ref{fig:setup}. At a
distance of about 6m from the interaction point in the ZEUS
calorimeter, a spaghetti type calorimeter, the so-called six meter
tagger (6mT), was installed. The magnetic field of the HERA magnet, in
which the 6mT was located, bent the low-angle scattered positrons to
the tagger, and thus the tagger was used for tagging photoproduction
events. The luminosity was determined by measuring the rate of photons
from the Bethe-Heitler (BH) process ($e^+ p \to e^+ \gamma p$). The
latter was measured by two independent systems, the photon calorimeter
(PCAL) and the spectrometer (SPEC). The two components (PCAL and SPEC)
enabled the measurement of the luminosity in two independent ways with
1\% relative uncertainty. Overall there was a 2.6\% uncertainty common
to both, due to the geometrical acceptance of the exit window.

A dedicated trigger logic was designed to collect photoproduction
events and keep the rates at acceptable levels. Two conditions had to
be fulfilled, a low-angle scattering positron candidate detected in
the 6mT and some activity in the main detector.

For calculating the acceptance corrections in the main detector, the
{\sc pythia} 6.221~\cite{pythia} generator was used. It was coupled to
the {\sc heracles} 4.6~\cite{heracles} program to simulate
electromagnetic radiative effects. The acceptance of the 6mT is 100\%
for photoproduction events with positrons in the approximate energy
range 3.5 - 6.8 GeV, with a mean $W$ of 287 GeV for HER and 203 GeV
for LER.

\section{Results}

The data samples taken with the dedicated trigger configuration
consisted of 4,164,552 events (corresponding to a luminosity of 560.9
nb$^{-1}$ in the HER and 12,798,828 events (corresponding to a
luminosity of 994.7 nb$^{-1}$) in the LER. These samples included
background from different sources, like beam-gas events, BH overlays
where the high energy photon hit the PCAL and the positron hit the
6mT, and off-momentum (with some transverse momentum) beam
positrons. After all cuts needed for suppressing background events,
62,653 events in the HER and 113,362 events in the LER remained as
genuine photoproduction events.

In order to calculate the ratio $r$ from Eq.~(\ref{acc}), the
acceptances $A$ were calculated using the MC simulation. The
acceptances of HER and LER are equal within errors,
corroborating our hypothesis that the ratio is independent of the
acceptance and thus reduces the systematic error coming from the MC
simulation. The systematic errors on the ratio came from the background
suppression cuts (1.05\%), from the uncertainty on the ratio of
luminosities (1\%) and from the uncertainty on the ratio of fluxes
(3.5\%). This last high systematic uncertainty comes from the fact
that the y range over which the flux has to be integrated is presently
not precisely determined by the 6mT. This will improve with a better
understanding and calibration of the 6mT.

The value of $\delta$ as obtained from $r$ is
\begin{equation}
\delta = 0.140 \pm 0.014(\rm{stat.}) \pm 0.042(\rm{syst.}) \pm 0.100(\rm{6mT}),
\end{equation}
which translates into  
\begin{equation}
\epsilon = 0.070 \pm 0.007(\rm{stat.}) \pm 0.021(\rm{syst.}) \pm 0.050(\rm{6mT}),
\end{equation}
This result is consistent with earlier determinations of $\epsilon$,
however has the advantage of being obtained from a single experiment
and being model independent.

The statistical uncertainty will be improved in the future by
including the data taken with a third proton beam energy (MER). The
systematic uncertainty will improve by a better understanding and
calibration of the 6mT.

\begin{footnotesize}

\end{footnotesize}



\begin{thebibliography}{99}
\bibitem{DL} A.~Donnachie and P.V.~Landshoff, 
\pl {B 296} {1992} {227}.


\bibitem{ZEUSpaper92} ZEUS Collaboration, M.~Derrick et al., 
\pl {B 293} {1992} {465-477}.

\bibitem{H1paper93} H1 Collaboration, T.~Ahmed et al., 
\pl {B 299} {1993} {374}.

\bibitem{ZEUSpaper94} ZEUS Collaboration, M.~Derrick et al., 
\zp {C 63} {1994} {391-408}.

\bibitem{H1paper95} H1 Collaboration, S.~Aid et al., 
\zp {C 69} {1995} {27}.

\bibitem{ZEUSpaper02} ZEUS Collaboration, S.~Chekanov et al., 
\np {B 627} {2002} {3-28}.

\bibitem{Cudell} J.R. Cudell et al., 
\prev {D 61} {2000} {034019}.
\bibitem{cudell2} J.R. Cudell et al., 
\pl {B 395} {1997} {311}.

\bibitem{dataset} Data used were extracted from the PPDS accessible at http://wwwppds.ihep.su:8001/ppds.html. 
Computer readable data files are also available at http://pdg.lbl.gov.


\bibitem{pythia} T. Sj\"ostrand, \zp {C 42} {1989} 301;\ \ H-U. Bengtsson and 
T. Sj\"ostrand, \cpc {46} {1987} {43}.

\bibitem{heracles} A.~Kwiatkowski, H.~Spiesberger and H.-J.~M\"ohring, 
\cpc {69}{1992}{155};\ \ H.~Spiesberger, {``{\sc heracles}, An event generator 
for $ep$ interactions at HERA including radiative processes (Version
4.6)''} (1996), available on http://www.desy.de/{\small
$\sim$}hspiesb/heracles.html.
\end{thebibliography}
\end{document}